\documentclass[12pt,a4paper]{article}
\usepackage{subfigure}
\usepackage{graphicx}
\usepackage{color}
\usepackage{lipsum} 
\usepackage{fancyhdr}
\begin{document}
\title{\textbf{\large{Determination of the Stopping Power and Failure-time of Spacecraft Components due to Proton Interaction Using GOES 11 Acquisition Data\footnote{\copyright{International Journal of Engineering Science \& Technology (IJEST). ISSN:0975-5462, Vol. 3 Issue 8, pp 6532-6542, August 2011}}}}}
\vspace{0.2cm}
\author{{\normalsize{\bf Jibiri N. N.$^{1}$\footnote{Corresponding author}, Victor U. J. Nwankwo$^{1}$, and Michael Kio$^{2}$}}\\
\vspace{0.2cm}
{\normalsize{jibirinn@yahoo.com; victornwankwo@yahoo.com}}\\
{\normalsize{$^{1}$ Radiation and Health Physics Research Laboratory,}}\\
{\normalsize{ Department of Physics, University of Ibadan, Nigeria}}\\
{\normalsize{$^{2}$ National Space Research and Development Agency (NASRDA), Abuja, Nigeria}}}
\date{}
\maketitle
\begin{abstract}
One of the several ways to describe the net effect of charged-particles' interaction is the rate of energy loss along the particles' path. In this study, the mass stopping power (Sp) of selected spacecraft composite materials, through which the particle traverses, its range (R) and the distance (d) traveled (by the particles) through the materials have been calculated and analyzed. The dose (in Gy) as a function of particle flux and deposited energy was also determined. Predictions of their possible effects on space system operations and life-span were made, especially as values exceeded certain threshold (limit). Using GOES 11 acquired data for 3 months, estimations and/or calculations were made to determine the risk and safe period of a satellite in the geosynchronous orbit. Under certain space radiation environmental conditions (without mitigation of any sort), a spacecraft whose body is 20 mm thick and with Al alloy casing, was theoretically estimated to have a safe period of about 3 years and risk period of about 29 years (due to total ionizing dose) within which it would experience a catastrophic failure.\\

\end{abstract}

\section{\large{Introduction}}
The Space is the void that exists beyond any celestial body including the earth. It is not completely empty but, in reality, the space is both a complex and a dynamic place[1] that is filled with energetic particles, radiation, and trillions of objects both very large and very small. Compared to what we experience on earth, it is a place of extremes. Distances are vast. Velocities can range from zero to the speed of light. Temperatures on the sunny side of an object can be very high, yet extremely low on the shady side, just a short distance away. Charged particles continually bombard exposed surfaces. Some have so much energy that they pass completely through an object in space. Magnetic fields can be intense. The environment in space is constantly changing. All of these factors influence the design and operation of space systems. There is no formal definition of where space begins. International law based on a review of current treaties, conventions, agreements and tradition, defines the lower boundary of space as the lowest perigee attainable by an orbiting space vehicle. A specific altitude is not mentioned. By international law standard aircraft, missiles and rockets flying over a country are considered to be in its national airspace, regardless of altitude. Orbiting spacecrafts are considered to be in space, regardless of altitude [1]. Space radiation consists primarily of ionizing radiation which exists in the form of light-energy and charged particles in space. Sources of radiation in Earth space are categorized into four and include Plasma, Trapped Particles in the earth’s magnetic field, Solar Particles Events (SPE), Galactic Cosmic Rays (GCR).\\

Radiation in space is generated by particles emitted from a variety of sources both within and beyond our solar system [13]. These particles (mostly of ionizing radiation) in motion possess enormous energy and can completely pass through an object in space. Spacecraft (including satellites) are subject to bombardment by these nuclear particles and electromagnetic radiations from both external and on-board sources [12]. When they penetrate the surfaces of these space vehicles, the electrical, electronic and electro-chemical components in them may be affected in one form or the other. Among other effects radiation exposure may cause include (i) induction of sporadic and unexplainable errors in sensitive parts in spacecrafts (ii) degrade the critical properties of structural materials (iii) jeopardize flight worthiness of the spacecraft (iv) lead to catastrophic failure and possibly mission ending effect and (v) constitute transient and terminal health hazard to both on-board passengers and astronauts. This research will verify and address these challenges and offer recipe for successful space mission.\\

In the last 25 years the National Geophysical Data Centre (located in Boulder Colorado and a part of US Dept of Commerce, NOAA) recorded over 4500 spacecraft anomalies or malfunctions that have been traced to the effects of the space radiation environment [3]. A reset occurred in Hubble Telescope after upgrade in 1996, when the spacecraft flew through South Atlantic Anomaly, a region in space where protons are trapped in earth’s magnetic field. All parts of Galileo, a space satellite were subjected to thorough radiation testing after its failure. It was discovered that failure did not occur until radiation level was close to design level [4]. These incidents and instances of spacecrafts anomalies and/or failures among host of others, is a challenge and a threat to space technology and calls for attention of researchers and hence a justification for this research. The aims and objectives of this study include an overview of (i) natural space radiation environment (ii) composition and intensity of space radiation (iii) effects of space radiation on life, electrical, electronic and electro-chemical components in spacecrafts. It will further compare materials used in constructing spacecraft and its Electrical, Electronics and Electrochemical Components (EEEC) in order as to assist designers and manufacturers in selection of materials for better design and proper implementation to achieve maximum system efficiency  and finally suggest ways of significantly lowering radiation effects and thus system failure time. This will save program cost and resources, through estimations and calculations using energy flux values from GOES 11 satellite in the geosynchronous orbit.\\

\section{Materials and Method}
\subsection{Collection of Data}
Secondary data for solar particle flux and energy in space was collected. This data was primarily collected by a Satellite Geostationary Operational Environmental Satellite (GOES -11), and prepared by the U.S. Department of Commerce, NOAA, Space Weather Prediction Center. GOES 11 is an American weather satellite, which is part of the US National Oceanic and Atmospheric Administration's Geostationary Operational Environmental satellite system. It was launched in 2000, and as at 2009 it was operating at the GOES-WEST position, providing coverage of the west of the United States. The data consist of proton flux ($E=1-100MeV$) and electron flux ($E=0.8-2MeV$) in GEO, taken at five minutes interval on daily basis for the period of three months (April –June, 2010).

\subsection{Interpretation of Data used}
The Label used for particles flux has the following interpretation: $P>1$ = Particles at $>1MeV$, $P>5$ = Particles at $>5MeV$, $P>10$ = Particles at $>10MeV$, $P>30$ = Particles at $>30MeV$, $P>50$ = Particles at $>50MeV$, $P>100$ = Particles at $>100MeV$, $E>0.6$ = Electrons at $>0.6MeV$, $E>2.0$ = Electrons at $>2.0MeV$. Units: Particles = $Protons/cm^2-s-sr$, Electrons = $Electrons/cm^2-s-sr$

\subsection{Analytical Procedures}
The mean flux per day was determined for each month; they were compared by means of graphs and logical interpretations. The composite materials of Spacecraft and its electronic components were considered for assessment and analysis (Germanium, Silicon, Aluminum and Aluminum alloy). The constituent elements of the Aluminum alloy considered and their composition (by weight) are presented in Table 1. The mass stopping power of the materials for the particle (with E=1-100MeV) was calculated as well as the Range and distance traveled by the particle as it traverses the materials. The Dose as a function of particle’s flux and energy was also calculated. Consequently, predictions of possible effects (of solar particles) on space system operations and life-span were made, particularly as values exceeded certain threshold or limit.

\begin{figure}
 \begin{center}
\includegraphics[height=4cm,width=10cm]{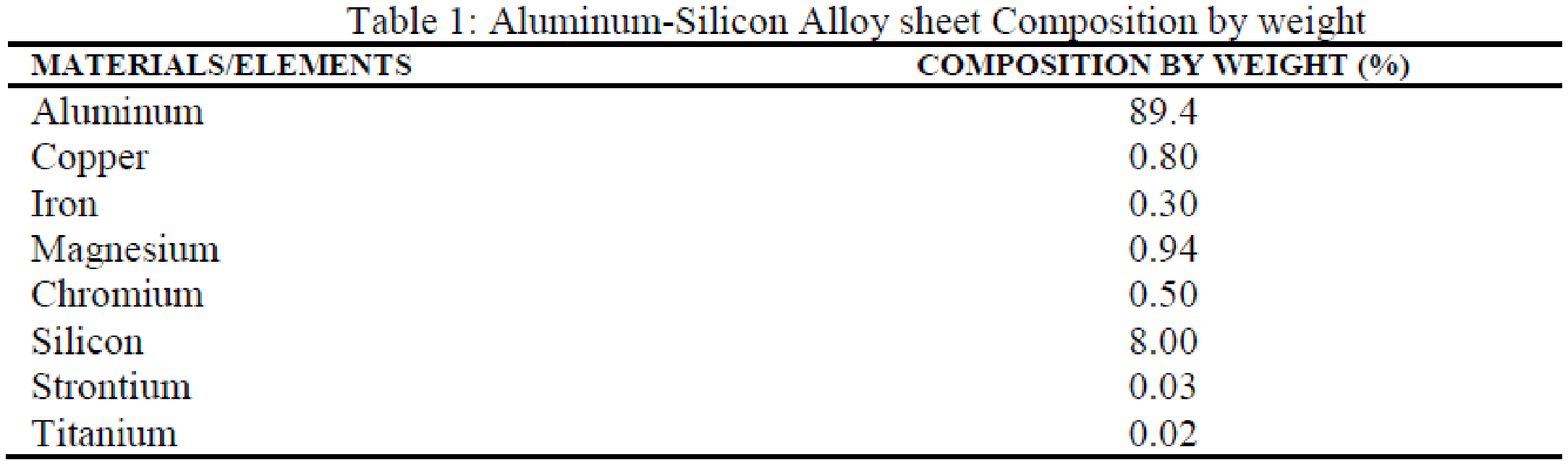}
\caption{Aluminum-silicon alloy composition by weight}
 \end{center}
\end{figure}

A more fundamental way of describing the rate of energy loss is to specify the rate in terms of the density thickness; this is called the Mass Stopping Power and has the unit $MeVcm^2/g$. The empirical relation below was adopted [5].

 \begin{eqnarray}
 -\frac{dE}{\rho dx} = \frac{a}{A} E^{-b}Z^{clogE+d} 
 \end{eqnarray}
The appropriate values of the constants and are a=915, b=0.85, c=0.145, d=0.635\\

The combined mass stopping power of the Aluminum-Silicon Alloy material in this work was calculated from the expression below:

 \begin{eqnarray}
\nonumber \left(-\frac{dE}{\rho dx}\right)_{combined} = \sum \%\ composition \times \left(-\frac{dE}{\rho dx}\right)_i of constituent\ elements 
 \end{eqnarray}
 
The dose (in Gy) as a function of the particle flux is also important, particularly as the spacecraft spends more time in the space radiation environment. The stopping power is used to determine dose from charged particle by the relationship:

 \begin{eqnarray}
\nonumber D = \phi \frac{dE}{\rho dx} 
 \end{eqnarray}
Where $\phi$ is the particle fluence (the number of particle striking the material over a specified time interval), its unit is $cm^{-2}$. D is measured in MeV/g. converting this to units of dose in the relation we get;

 \begin{eqnarray}
 D = \phi \frac{dE}{\rho dx}(1.6 \times 10^{-10}) Gy 
 \end{eqnarray}
Where 1Gy = 100rad

The Range, R of a proton with initial kinetic energy $E_\circ$, mass m, is mean distance it traveled before it stops. The range depends upon the type of the particle, its initial energy and the material it traverses. It is expressed in the unit of $g/cm^2$ or $mg/cm^2$. We adopted the empirical relation [5] below;

\begin{eqnarray}
 R_p = m_pG_pE^{1.85}Z^{-0.145logE} + F_p
\end{eqnarray}

\begin{eqnarray}
\nonumber G_p = \left(\frac{A}{915 \times 1.85Z^{0.635} \left(1 - \frac{0.145logZ}{1.8}\right)}\right)
\end{eqnarray}

\begin{eqnarray}
\nonumber F_p = R_1(E_1) - m_pG_pE^{1.85}Z^{-0.145logE_1}
 \end{eqnarray}
The Range is related to the distance, x traveled by the particle (in cm) by the equation:
 \begin{eqnarray}
 R = \rho x 
 \end{eqnarray}
Where $\rho$ is the density of the material through which the particle traverses.

\subsubsection*{Limitations of equations used}
The equations adopted for calculations in this work, particularly the mass stopping power (1) was originally obtained and used for calculation of mass stopping power of low energy region. By way of comparison calculations made using Bethe’s equation showed that at higher energies ($E>50MeV$) the difference between equation (1) and Bethe’s equation becomes significant.

\section{Theoretical Calculation/Prediction of Spacecraft Failure Time}
Satellite and Space Probes typically encounter Total Ionizing Dose between 10 (100Gy) and 100 krad (Si) (1000Gy). A spacecraft made of Al alloy (20mm thick), housing sensitive EEEC in the absence of mitigation of any kind becomes susceptible when particles with $E \geq 78MeV$ bombards it. Greater percentage of this energy is lost during the process due to the stopping power of the alloy; however the remainder constitutes a significant dose to the EEEC components. The cumulative equivalent dose from this energy spectrum is about 259 rad(Si) in Silicon and 120 rad(Ge) in Germanium. The net effect is that deposited dose builds up with time as exposure continues until the threshold is exceeded and consequently failure occurs due to Total Ionizing Dose (TID). To predict the Mean Time of Failure, the time to failure t (in years) is given by the suggested equation:

 \begin{eqnarray}
 t_{yrs} = \frac{TID_{threshold}}{Dose/yr} 
 \end{eqnarray}
\
\begin{figure}[h]
 \begin{center}
 \includegraphics[height=4cm,width=10cm]{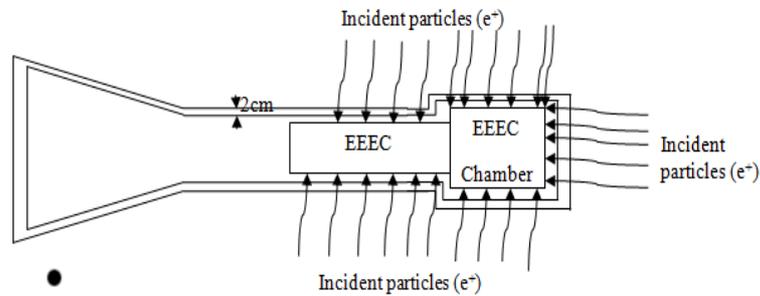}
 \caption{Spacecraft exposure to energetic particles in space environment}
 \end{center}
\end{figure}
\
\section{Results and discussion}
\subsection*{Calculation of Parameters}
Using the empirical relations in equations 1, 3 and 4, the mass stopping power and particle's Range and distance in Aluminum, Aluminum Alloy, Germanium and Silicon were calculated for proton energy of E=1-100MeV. The results are presented in Tables 2 - 5. The dose deposited by particles as it traverses through the materials (Ge, Si) was also calculated from equation 2 and presented in Tables 6 and 7. The mean particles (Electron and Proton) fluxes per day were determined for each month; they were compared by means of graphs and logical interpretations. Aluminum alloy material gave increased stopping power compared to pure Aluminum.  This implies that Al alloy may be a better shield from particles than pure Al (see Figure 3 and Table 2). As could be seen Ge is more prone to particle penetration than Si. Though dose deposited by the particle at same energy is much less in Ge compared to Si, Ge receives twice as Si (Figure 3 and Table 2). Between 2nd and 11th day, electron flux tend to increase with the peak occurring in April (Figure 4). There is a possibility of the spacecraft being in eclipse/shadow between 2nd and 11th. The Satellite/Spacecraft (in GEO) may have experienced negative charging and consequently assume a negative potential. Proton flux ($E \geq 10MeV$) climaxed in June between 10 – 13th (Figures 5 and 6). In addition to trapped radiation in the geostationary orbit a solar particle event (such as CMEs, solar flares) may have occurred leading to an increase in particle flux. This implies increased dose to spacecraft components in the event of particles access to the components. In addition, the probability of Single-event effects occurrence in EEEC is most likely. Aluminum alloy should be used in building the body of spacecrafts. Where the recommendation for the use of Aluminum has been implemented, increase in percentage composition of constituent elements with high stopping power (such as Mg and Si) should also be considered. However, this should be carefully done so as to avert inherent problem resulting from properties of individual constituent elements such as expansion and contraction during temperature rise/fall, and associated weight of the craft. A multi-layer surface coating should be considered in addition to using materials with high stopping power such as Magnesium, Silicon, Carbon and Beryllium.\\

\begin{figure}[h]
 \begin{center}
\includegraphics[height=6cm,width=10cm]{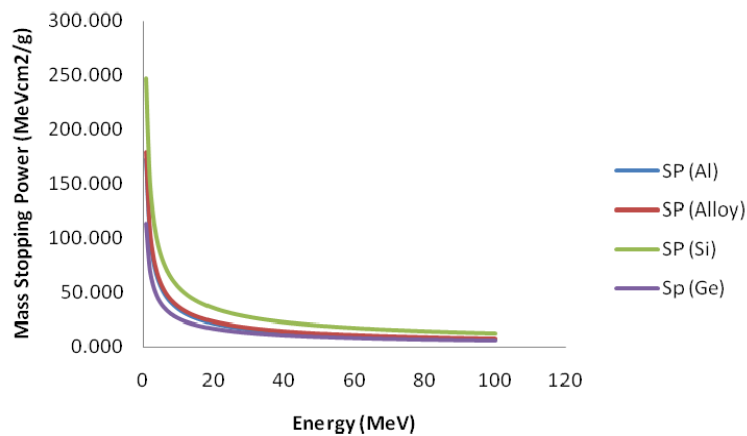}
\caption{Combined Plot of mass stopping power (Al, Al alloy, Ge and Si)}
 \end{center}
\end{figure}
\begin{figure}[h]
 \begin{center}
\includegraphics[height=6cm,width=12cm]{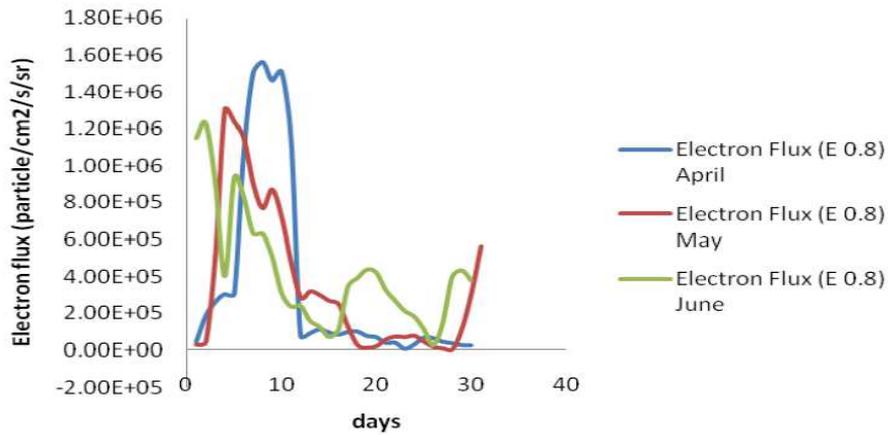}
\caption{Electron flux (E 0.8) against days of the month (April, May and June)}
 \end{center}
\end{figure}
\begin{figure}[h]
 \begin{center}
\includegraphics[height=6cm,width=10cm]{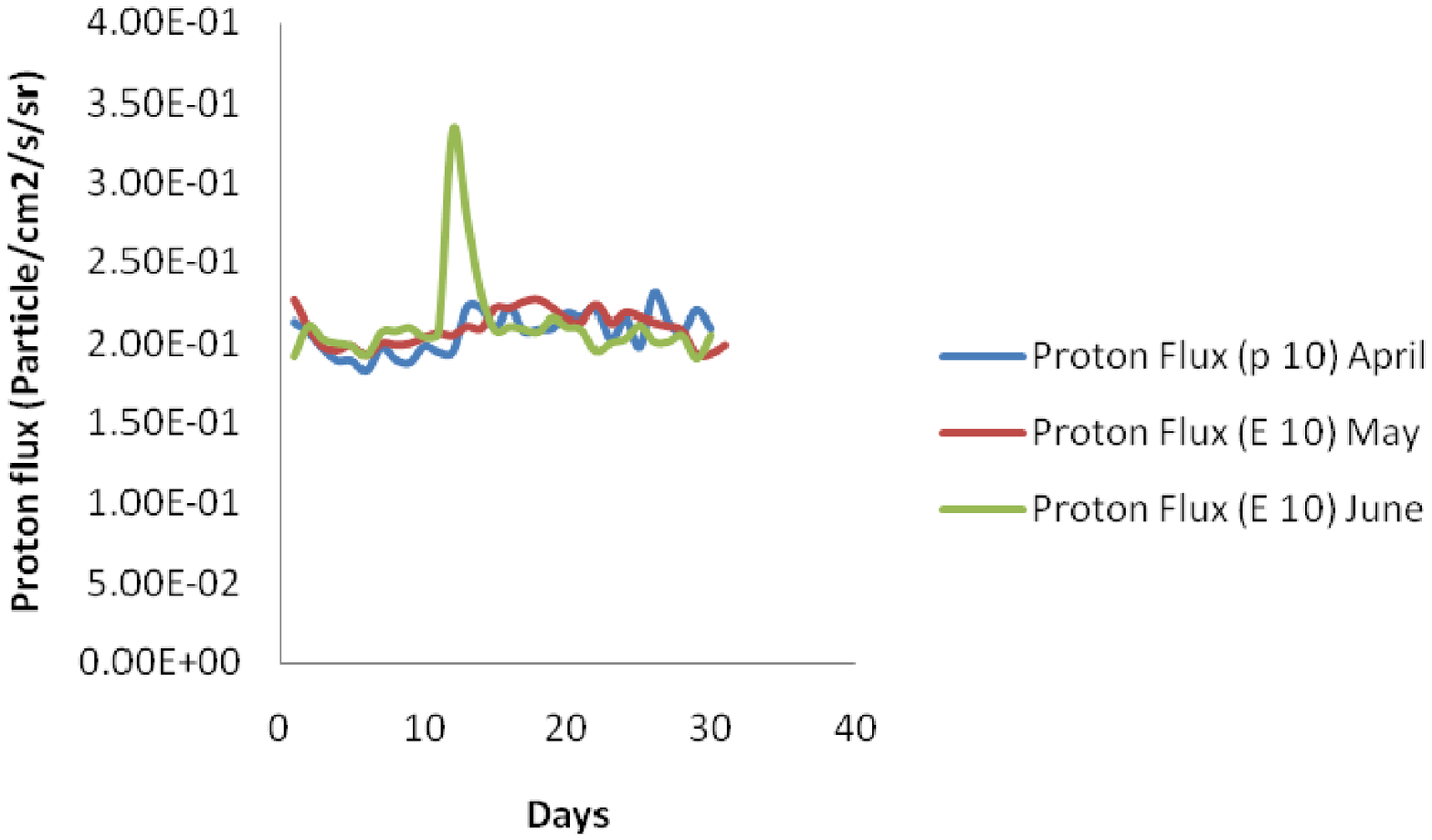}
\caption{Proton flux (P 10) against days of the months (April, May, June)}
 \end{center}
\end{figure}
\begin{figure}[h]
 \begin{center}
\includegraphics[height=6cm,width=10cm]{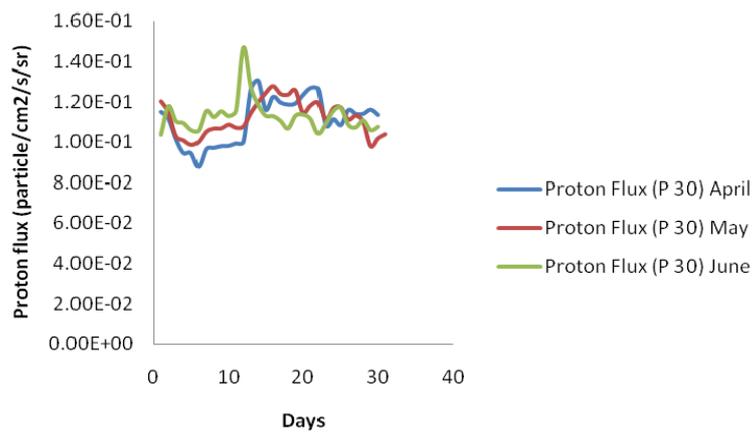}
\caption{Proton flux (P 30) against days of the month (April, May, June)}
 \end{center}
\end{figure}
\clearpage
\newpage

\section{Conclusion}
Aluminum alloy may better shield the spacecraft and its EEEC against particle than pure Aluminum because it high stopping power. Electron flux was at peak between 2nd and 11th of April 2010 considering the three months data. This could result to a negative potential build up on spacecraft surface (depending on it position), of which upon emergence into sunlight could experience a possible discharge effects which can disrupt satellite operations. Proton flux peaked between 10th and 13th of June, an indication that a solar particle event may have occurred. This exposes the space system to high dose rate and susceptibility to Single event Effects. Under these space radiation environmental conditions, in the Geosynchronous orbit and in the absence of mitigation of any kind a spacecraft whose body is made of Al alloy and 20mm thick, housing sensitive EEEC will theoretically experience minimum failure-time threshold after 3 years (10 krad) and continue to be at risk until 29th year when dose build-up equals 100krad (due to Total Ionizing Dose), during which it will experience a catastrophic failure.
\
\section*{Acknowledgments}
With a grateful heart we wish to acknowledge Dr. Godstime James of the National Space Research and Development Agency (NARSDA) Abuja, Nigeria, the agency itself and Physics Department, University of Ibadan for the technical assistance during this work.

\begin{figure}[h]
 \begin{center}
\includegraphics[height=18cm,width=12cm]{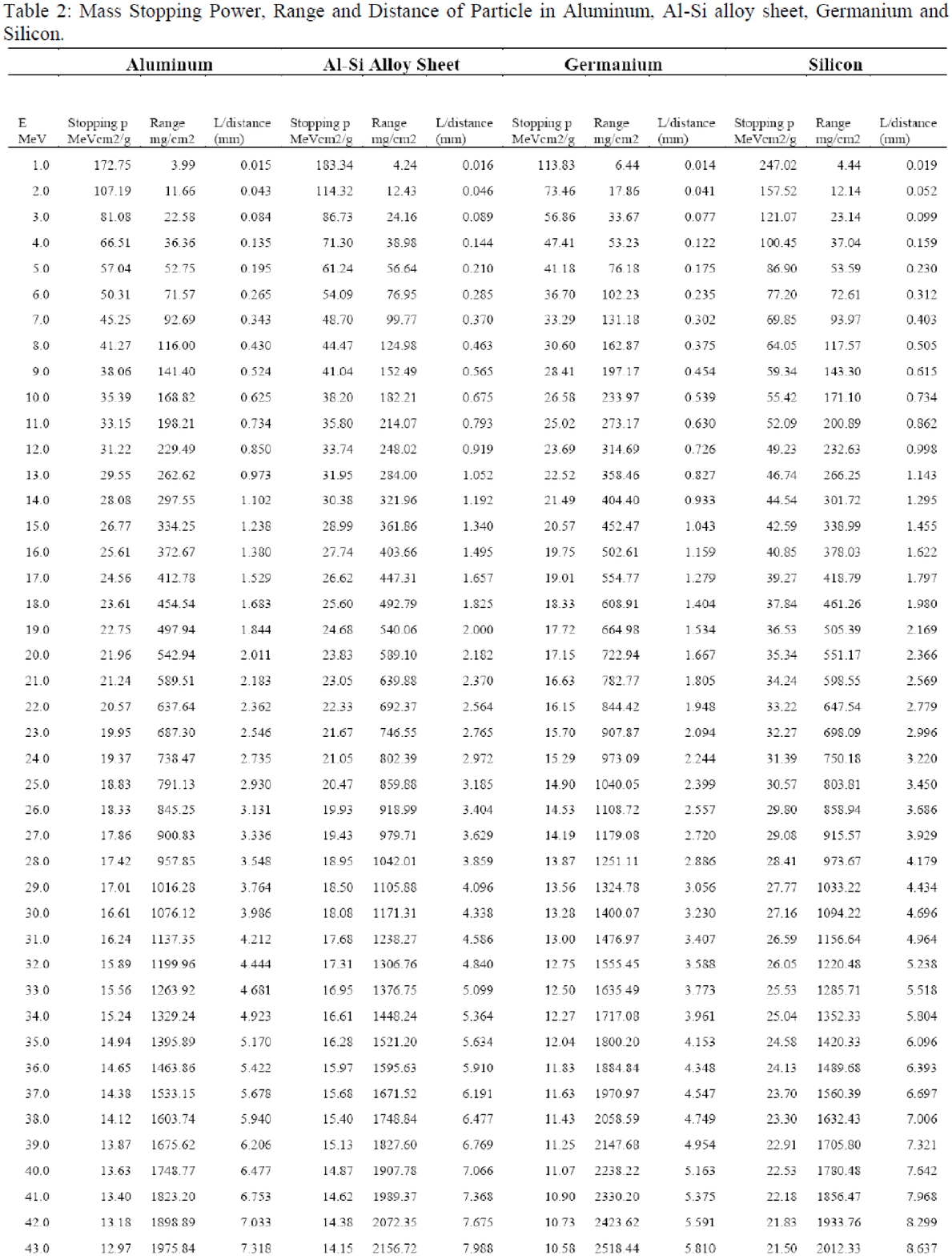}
\caption{Mass stopping power, Range and distance of particle in aluminum, Al-Si alloy sheet, Germanium and Silicon}
 \end{center}
\end{figure}
\
\begin{figure}[h]
\includegraphics[height=18cm,width=12cm]{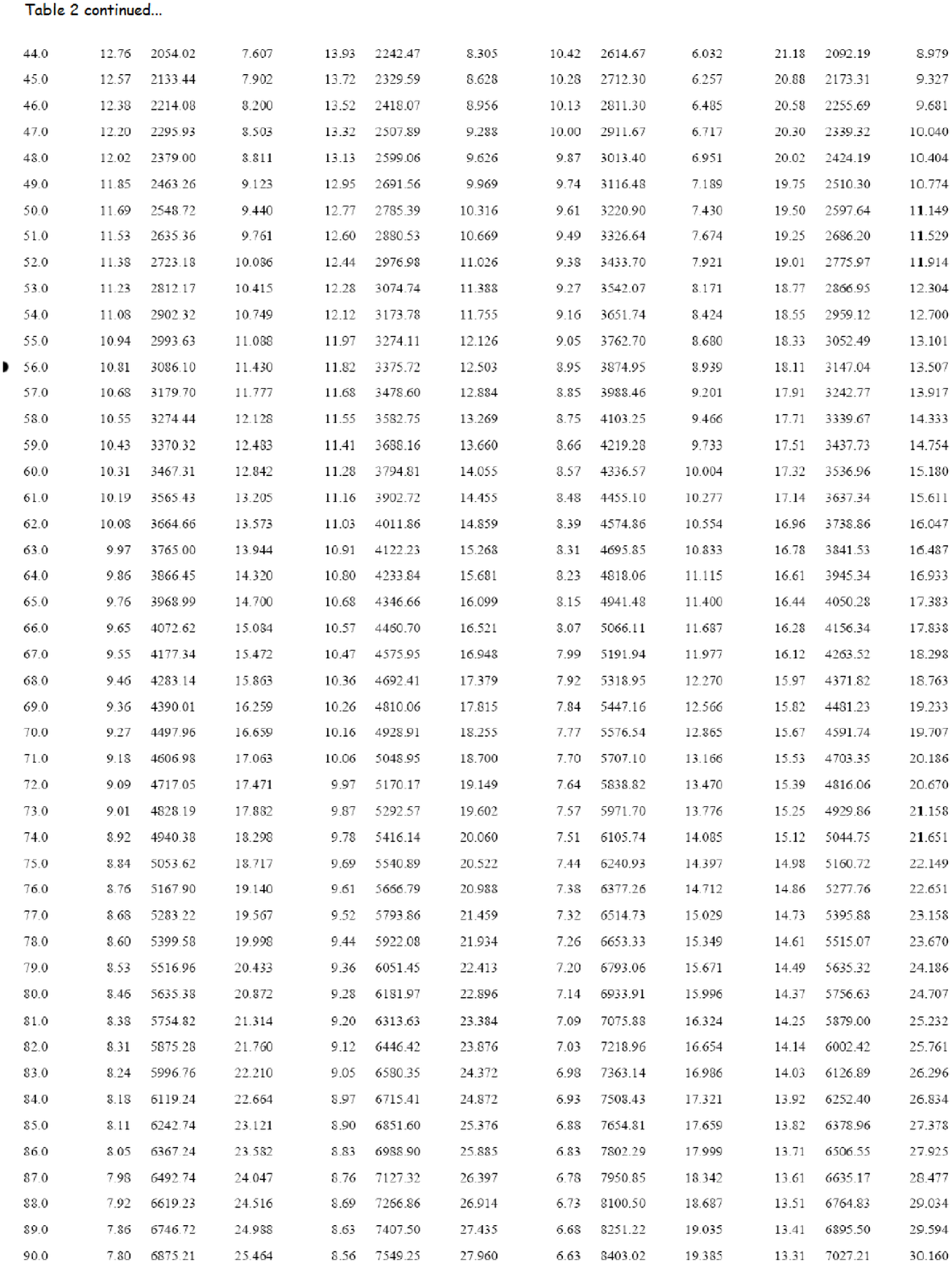}
\caption{Mass stopping power, Range and distance of particle in aluminum, Al-Si alloy sheet, Germanium and Silicon}
\end{figure}
\
\begin{figure}[h]
 \begin{center}
\includegraphics[height=5cm,width=12cm]{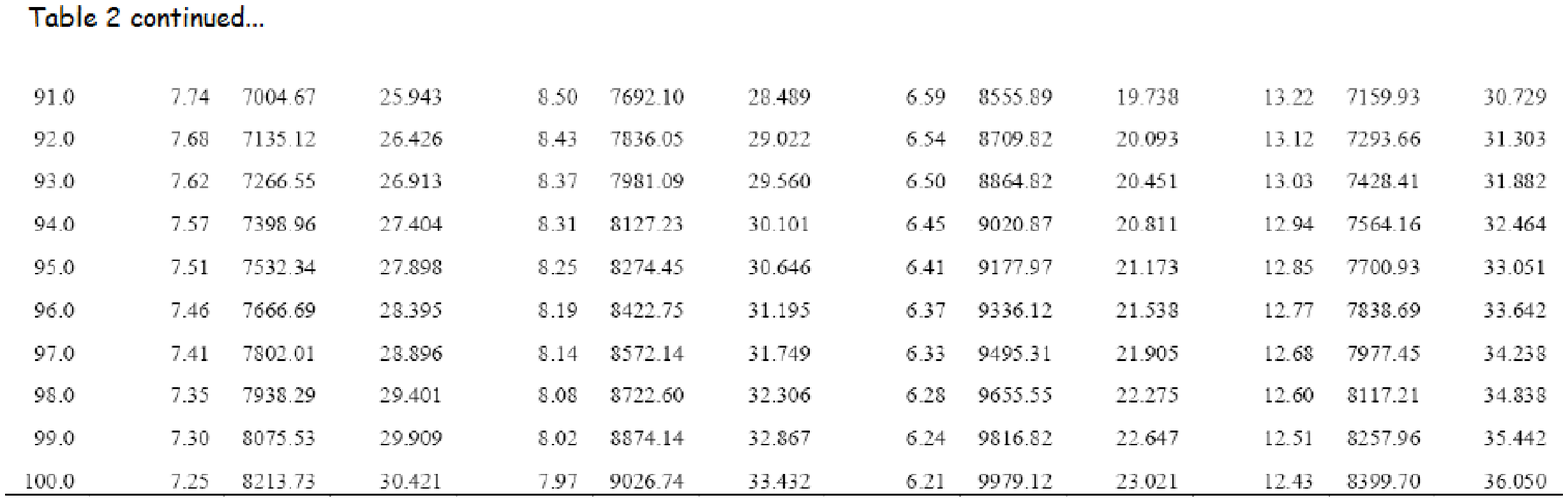}
\caption{Mass stopping power, Range and distance of particle in aluminum, Al-Si alloy sheet, Germanium and Silicon}
 \end{center}
\end{figure}
\
\begin{figure}[h]
 \begin{center}
\includegraphics[height=10cm,width=12cm]{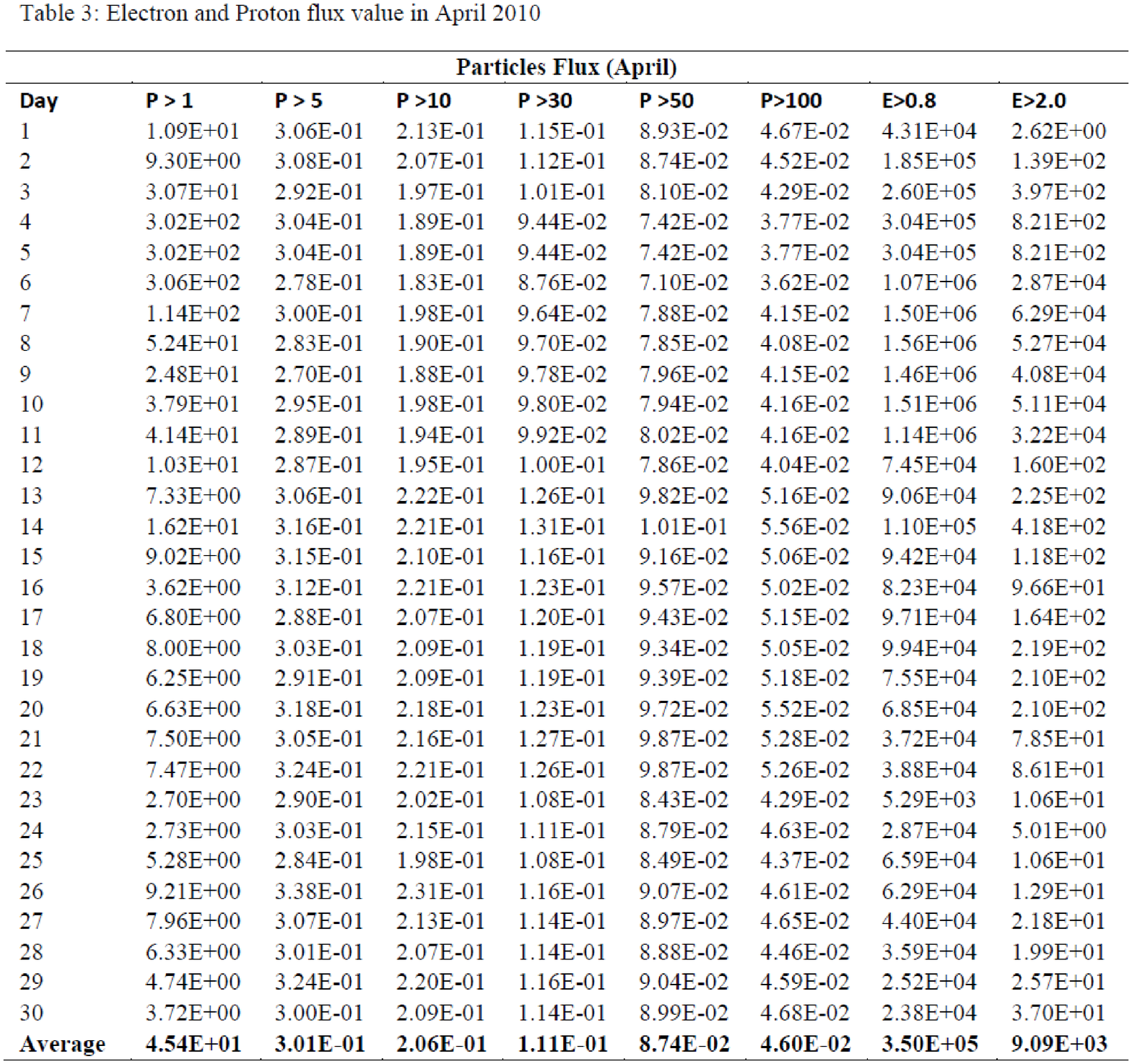}
\caption{Electron and Proton flux in April}
 \end{center}
\end{figure}
\
\begin{figure}[h]
 \begin{center}
\includegraphics[height=10cm,width=12cm]{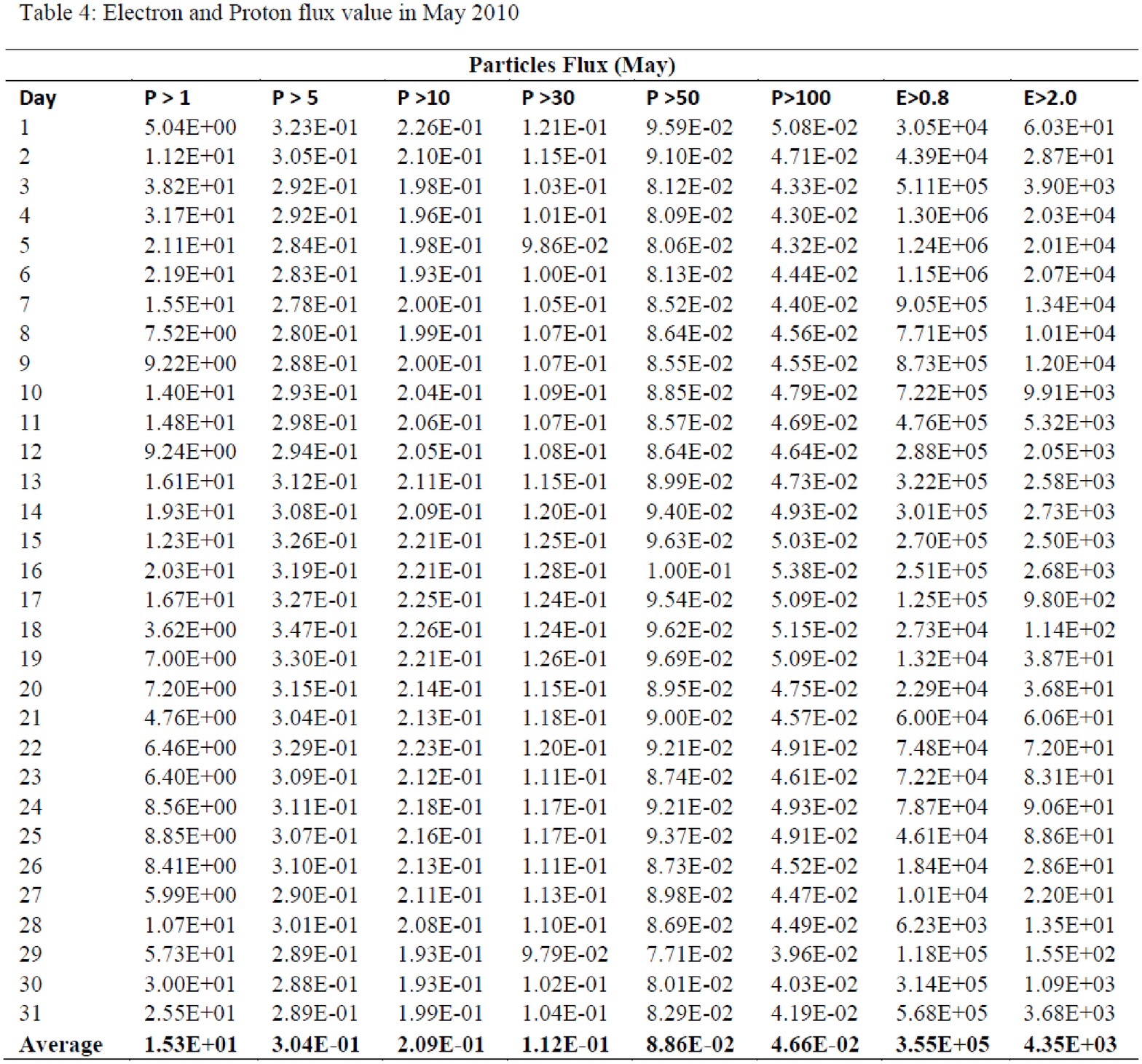}
\caption{Electron and Proton flux in May}
 \end{center}
\end{figure}
\
\begin{figure}[h]
 \begin{center}
\includegraphics[height=8cm,width=12cm]{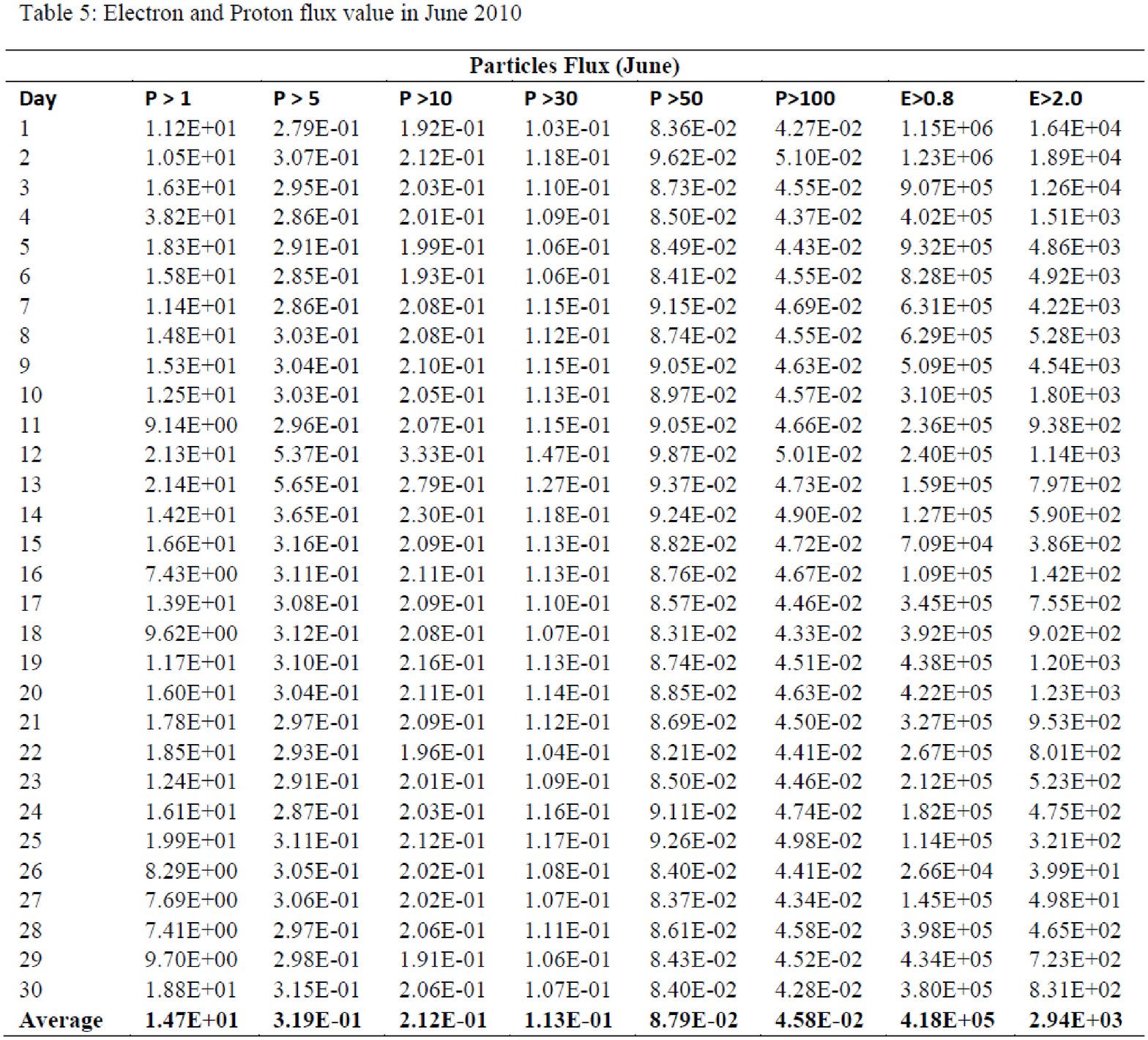}
\caption{Electron and Proton flux in June}
 \end{center}
\end{figure}
\
\begin{figure}[h]
 \begin{center}
\includegraphics[height=4cm,width=12cm]{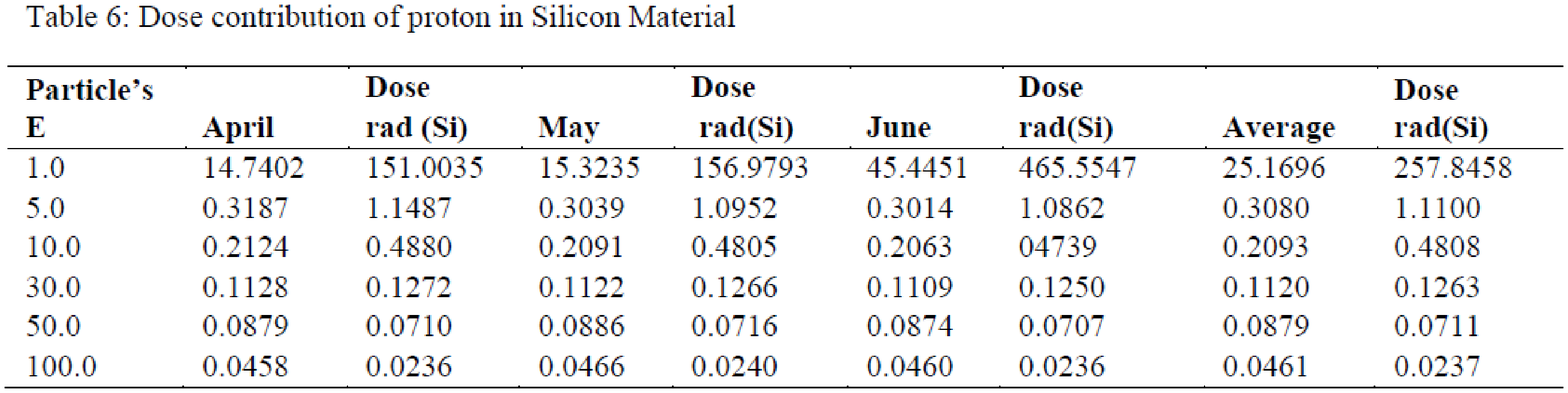}
\caption{Dose contribution of proton in Silicon Material}
 \end{center}
\end{figure}
\
\begin{figure}[h]
 \begin{center}
\includegraphics[height=6cm,width=12cm]{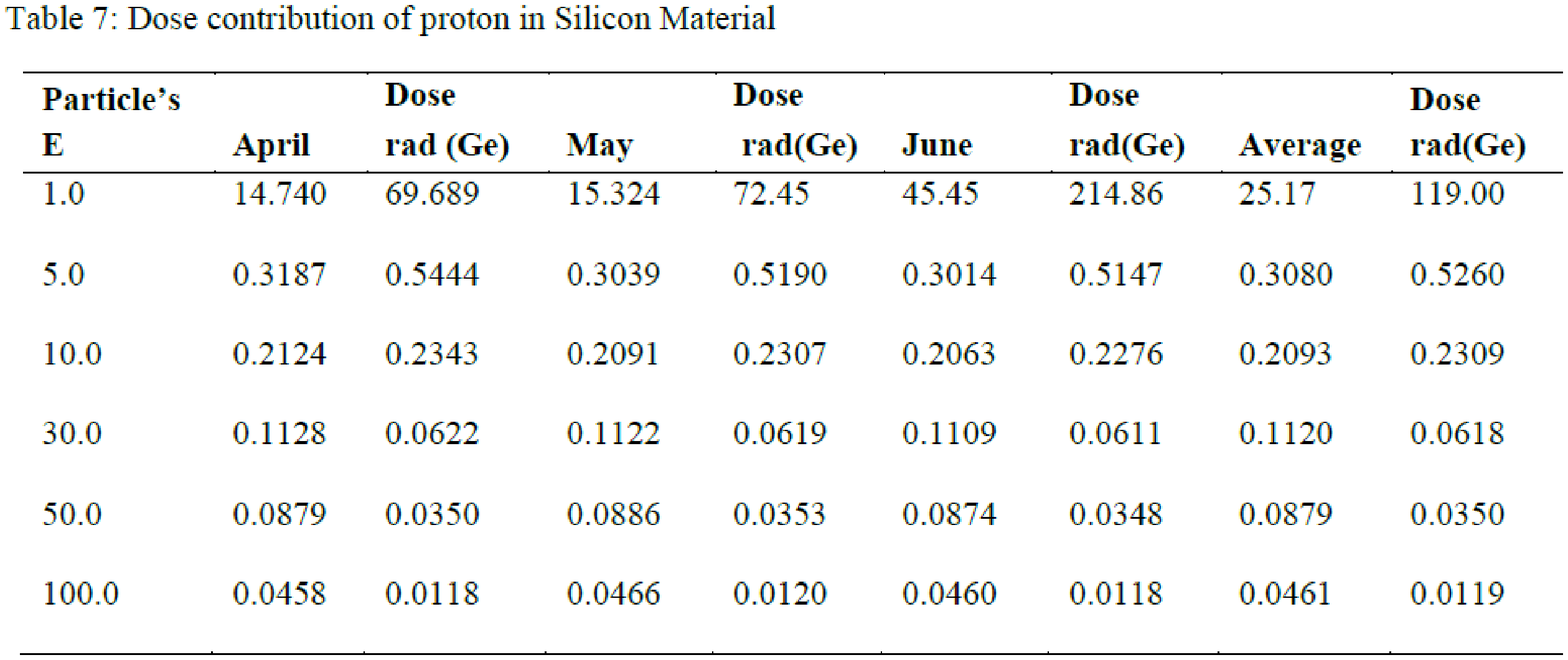}
\caption{Dose contribution of proton in Germanium Material}
 \end{center}
\end{figure}
\end{document}